\documentclass[%
reprint,superscriptaddress,
 amsmath,amssymb,
prl,
]{revtex4-1}

\usepackage{aas_macros}
\usepackage{graphicx,mathrsfs}
\usepackage{dcolumn}
\usepackage{bm}
\usepackage{color}
\definecolor{colorLink}{rgb}{0.9,0,0} 
\definecolor{colorCite}{rgb}{0,0.7,0} 
\definecolor{colorURL} {rgb}{0,0,0.8} 
\usepackage[colorlinks=true,linktocpage=true,linkcolor=colorLink,citecolor=colorCite,urlcolor=colorURL]{hyperref}

\newcommand{\HI}{H\textsc{i}\,\,}

\begin{document}

\title{Comment on the paper\\ ``Calorimetric Dark Matter Detection with Galactic Center Gas Clouds"}

\date{January 14, 2020}

\author{Glennys R. Farrar}
 \email{gf25@nyu.edu}
\affiliation{
Center for Cosmology and Particle Physics, Department of Physics, New York University, New York, NY 10003, USA
}
\author{Felix J. Lockman}
\email{jlockman@nrao.edu}
\affiliation{Green Bank Observatory, Green Bank, WV 24944, USA}

\author{N. M. McClure-Griffiths}
\email{naomi.mcclure-griffiths@anu.edu.au}
\affiliation{Research School of Astronomy \& Astrophysics, Australian National University, Canberra ACT 2611, Australia.}
\author{Digvijay Wadekar}
 \email{jay.wadekar@nyu.edu}
 \affiliation{
Center for Cosmology and Particle Physics, Department of Physics, New York University, New York, NY 10003, USA
}
 

\maketitle
In a recent Letter, Bhoonah \emph{et al}. \cite{BhoBramante18}
(hereafter B18) attempted to derive limits on dark matter interactions
with ordinary matter by demanding that DM heating of gas clouds not
exceed the known astrophysical cooling rate based on the temperature, density and metallicity of the observed clouds.   In B18, the cloud
G1.4$-$1.8+87 from \cite{McClureLock13} (hereafter McG13) was singled
out as most suitable by virtue of its apparently exceptionally low temperature
and relatively low density.  In this Comment, we point out a fundamental conceptual error in B18, namely their use of clouds in the high-velocity nuclear outflow (HVNO) of the Galaxy for the  analysis. 
This, along with additional detailed errors, invalidates the limits reported in B18.  

The conceptual error with B18 is their use of complex, poorly understood and likely-short-lived clouds for placing limits. 
The HVNO clouds are in the hot, high-velocity wind ($10^{6-7}$K, 330 km/s) emanating from the Galactic Center. This extreme environment causes shocks and other destructive effects, likely making the clouds transient objects \cite{cooper08_WindSim, ScaBru15_WindSim, SchRob17_WindSim, ArmFra17_WindSim,MelGou13_WindSim,McCourt18_WindSim,Gronke18_WindSim,Sparre18_WindSim}. However deriving DM bounds based on heat transport requires the system to be in a steady state at the current temperature over the long timescales associated with its purported radiative cooling rate, invalidating the use of a system for which the required stability is doubtful. The subsequent more detailed analysis in Bhoonah \emph{et al}. \cite{BhoBramante19} also ignores the effect of the extreme environment on the HVNO clouds and hence suffers from the same fundamental problem.  

\begin{figure}
\centering
\includegraphics[trim = 0.4in 0.1in 0.3in 0.2in,width=0.4\textwidth, keepaspectratio=true]{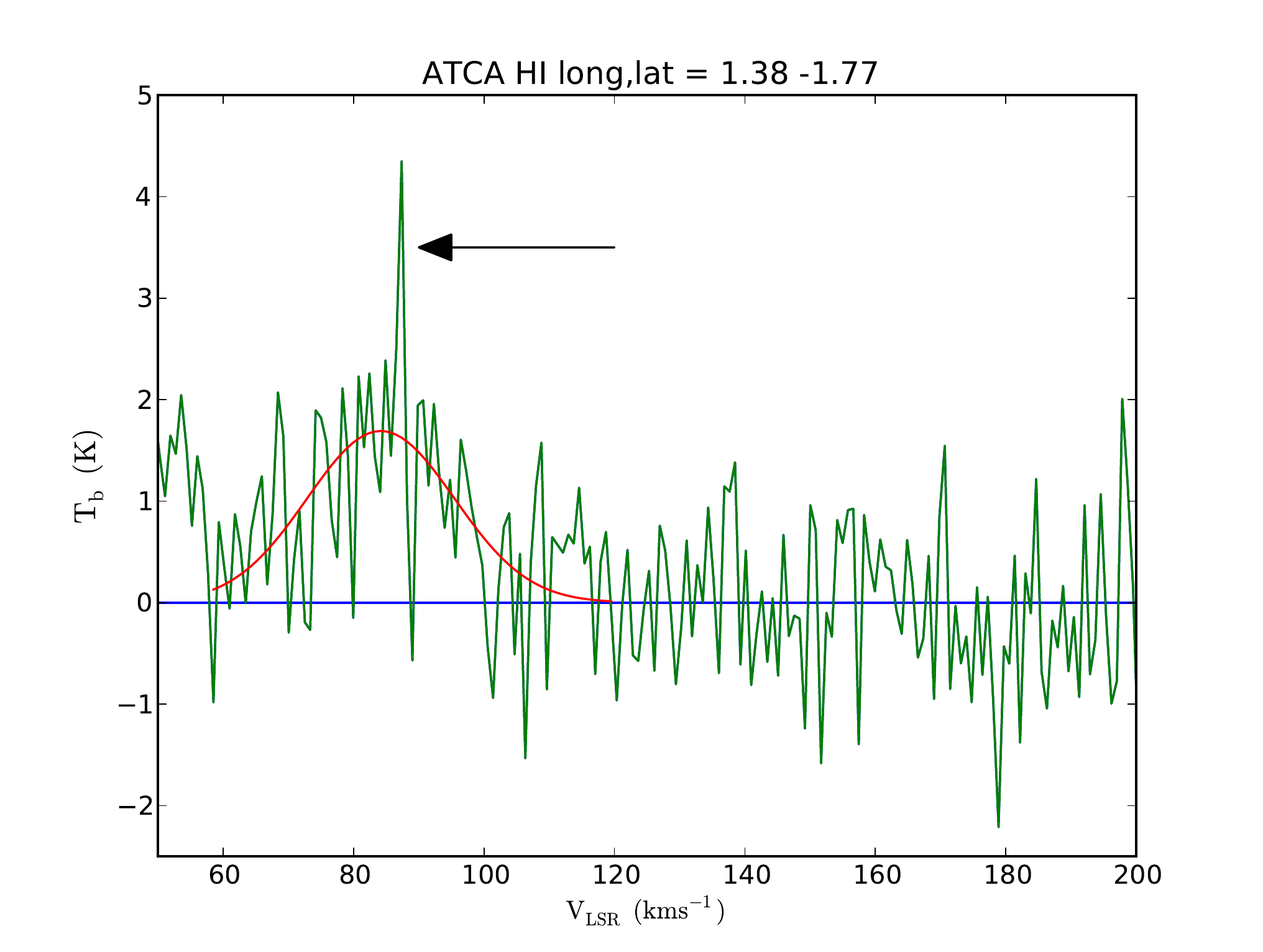}
\includegraphics[trim = 0.4in 0.1in 0.3in 0.1in,width=0.4\textwidth,keepaspectratio=true]{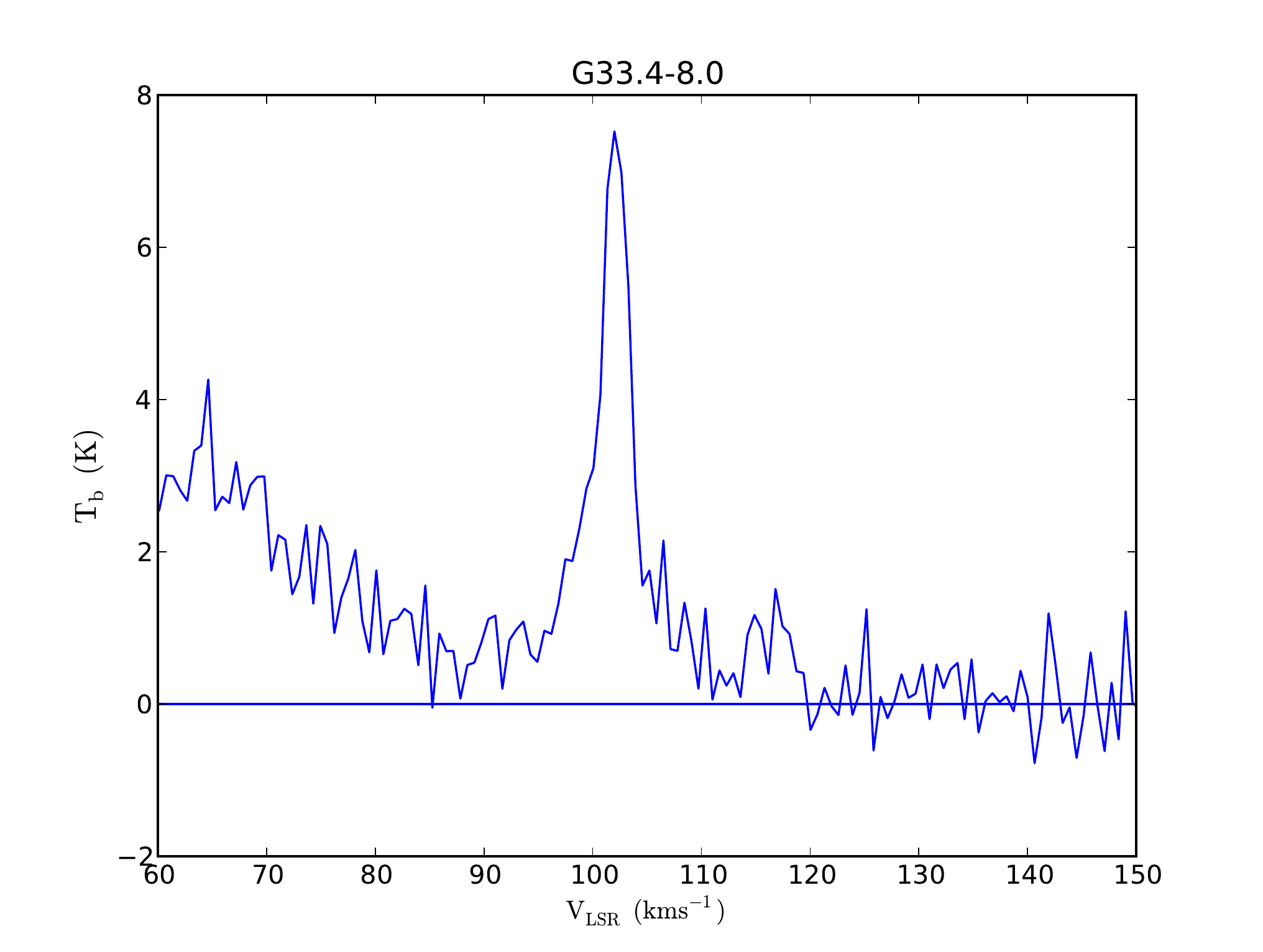}
\vspace{-0.15in}
\caption{{\it Top:} The \HI brightness temperature spectrum in the direction of G1.4-1.8+87. 
An arrow marks the extremely narrow line quoted in the McG13 table, while the
smooth curve shows a more appropriate Gaussian fit to the emission
feature. {\it Bottom:} The corresponding spectrum for G33.4-8.0 \cite{PidLock15}.}
\label{fig:Cloud} 
\vspace{-0.25in}
\end{figure}

A further problem is that B18 calculated the temperature of G1.4$-$1.8+87 to be $T_g \lessapprox 22$ K by taking the velocity dispersion to be 1 km/s.  Fig. \ref{fig:Cloud} shows the \HI\ spectrum at the
location of the cloud  G1.4$-$1.8+87, from the public online data \cite{MCGLock12}. As seen in Fig. \ref{fig:Cloud}, most of the \HI\ emission for this cloud is
characterized by a line with a FWHM of 26.6 km/s (red line), with the narrow 1 km/s spike being a single-channel fluctuation (see \cite{newObs}).   For comparison, the spectrum of a robust cloud G33.4-8.0 \cite{PidLock15}, used in \cite{wf19}, is also shown.

 Using the correct width 26.6 km/s gives $T_g $ above 15,000 K.  Some other parameters given in B18 for the cloud G1.4$-$1.8+87 are also in error: B18 quotes the mass and radius of cloud to be $311 M_\odot$ and 12 pc, whereas the correct values in McG13 are $17$ $M_\odot$ and 8.2 pc. The incorrect values of B18 appear to have been read from adjacent lines of the table in McG13.  The cooling function drops drastically for $T< 100$ K, so the net effect of correcting the temperature and the density is that the radiative cooling rate of G1.4$-$1.8+87 increases by a factor $\approx 10^6$ and thus the conclusions drawn by B18 from G1.4$-$1.8+87 are incorrect, even if using HVNO clouds were legitimate.
 
Two other errors in B18 need mentioning to avoid others follow their example.  First, B18 confuses the velocity of the cloud relative to the local standard of rest V$_\textup{LSR} = 87$ km/s, reported in  \cite{McClureLock13}, with the velocity of the cloud relative to the Galaxy's center of mass.  V$_\textup{LSR}$ is defined to be an object's line-of-sight velocity relative to a frame of reference in a circular orbit around the Galactic Center at the position of the solar system.  Instead, the velocity of the cloud relative to the Galaxy is to a good approximation the outflow velocity of the \HI clouds entrained in the nuclear wind, $\sim 330$ km/s from \cite{DiTeodoroLock18}. 

Second, a conservative bound requires adopting the smallest local DM density consistent with observations, which near the Galactic Center is generally given by the Burkert profile \cite{Burkert95}.  B18 takes incorrect parameter values which exaggerate the Burkert density by a factor of 9 (without citing a source), $ \rho_b = 14$ GeV/cm$^3$ and $r_b=3$ kpc,  instead of $\rho_b = 1.57$ GeV/cm$^3$ and $r_b=9.26$ kpc from the latest fit\cite{Nes13_BurkertFit}; the expression quoted in B18 for the form of the Burkert profile is also incorrect.

Limits on DM scattering from the cooling of suitable Milky Way clouds, and new and complementary constraints on DM from the Leo T dwarf galaxy, are reported in \cite{wf19}; a more detailed discussion of HVNO clouds is given in its Supplemental Material.   The true limits from Galactic clouds are $10^2 $ and $10^3$ times less stringent for the millicharge parameter $\epsilon$ and the DM-nucleon scattering cross section, respectively, than claimed in B18 (see \cite{wf19}). 

We thank C. McKee for helpful comments. GRF acknowledges support of NSF-1517319.  The Green Bank Observatory is a facility of the National Science
Foundation operated under a cooperative agreement by Associated
Universities, Inc.

\bibliographystyle{apsrev4-1}
%

\end{document}